\documentclass[english,twocolumn,showpacs,preprintnumbers,groupedaddress,amsmath,amssymb,aps,prl]{revtex4}
\usepackage[T1]{fontenc}
\usepackage[latin1]{inputenc}
\usepackage{textcomp}
\usepackage{graphicx}

\makeatletter
\usepackage{bm}
\makeatother
\usepackage{babel}

\begin{document}

\title{Optical precursors in transparent media}

\author{Bruno Macke}

\author{Bernard S\'{e}gard}

\email{bernard.segard@univ-lille1.fr}

\affiliation{Laboratoire de Physique des Lasers, Atomes et Molécules (PhLAM),
Centre d'Etudes et de Recherches Lasers et Applications, CNRS et Universit\'{e}
Lille 1, 59655 Villeneuve d'Ascq, France}

\date{\today}

\begin{abstract}
We theoretically study the linear propagation of a stepwise pulse
through a dilute dispersive medium when the frequency of the optical
carrier coincides with the center of a natural or electromagnetically
induced transparency window of the medium (slow-light systems). We
obtain \emph{fully analytical expressions} of the entirety of the
step response and show that, for parameters representative of real
experiments, Sommerfeld-Brillouin precursors, main field and second
precursors (\textquotedblleft{}postcursors\textquotedblright{}) can
be distinctly observed, all with amplitudes comparable to that of
the incident step. This behavior strongly contrasts with that of the
systems generally considered up to now.
\end{abstract}

\pacs{42.25.Hz, 42.25.Kb, 42.25.Lc}

\maketitle
As far back as 1914, Sommerfeld and Brillouin theoretically studied
the propagation of a stepwise pulse through a linear dispersive medium
\cite{som14,bri14,re0}. They showed in particular \cite{bri14}
that the arrival of the main signal is preceded by that of two successive
transients they named forerunners. The first one (now called the Sommerfeld
precursor) arrives with the velocity $c$ of light in vacuum. Its
instantaneous frequency, initially higher than the frequency $\omega_{C}$
of the optical carrier, decreases as a function of time whereas that
of the second one (the Brillouin precursor), initially lower than
$\omega_{C}$, evolves in the opposite direction. Sommerfeld and Brillouin
considered a single-resonance Lorentz medium and made their calculation
by using the newly developed saddle-point method of integration. Revisited by various methods, this problem has become a canonical problem in electromagnetics and optics \cite{stra41,jack75,oug94}.
Different models of medium have obviously been considered and the theoretical literature
on precursors is very abundant. See \cite{oug07} for a recent review. 

As intuitively expected, the precursors will be observed only if the
rise-time of the incident step is short compared to the response time
of the medium \cite{oug95}. Most of the theoretical papers consider
dense media with very short response time ($<1$ fs) and the fulfillment
of the previous condition raises serious experimental difficulties.
This explains the dramatic dearth of papers reporting direct demonstrations
of precursors. A first experiment was achieved in the microwave region
with waveguides whose dispersion mimics that of the Lorentz medium
\cite{ples69}. In the optical domain, Aavikssoo \emph{et al}. studied
the propagation of single-ended exponential pulses through a GaAs crystal \cite{aa91}. Associated with an exciton line, the precursors then appear as a spike superimposed on the main pulse (see also \cite{sa02}). A discussion on the observability of optical precursors in dense media can be found in \cite{ost07}. 

Much more favorable time scales are obtained by exploiting the narrowness
of atomic or molecular lines in vapors or gases. The switching times
of the incident field may then be very long compared to the optical
period without washing out the transients. In such conditions, the
slowly varying envelope approximation (SVEA) is absolutely justified. The medium is fully characterized by its system function $H(\Omega)$ connecting the Fourier transforms of the envelopes of
the transmitted and incident fields \cite{pap87}. $\Omega$ designates
the deviation of the current optical frequency $\omega$ from the carrier frequency $\omega_{C}$
and the envelope of the optical step response reads 
\begin{equation}
a(t)=\int_{\Gamma}H(\Omega)\exp\left(i\Omega t\right)d\Omega/2i\pi\Omega\label{eq1}
\end{equation} 
where the contour $\Gamma$ is a straight line parallel to the real
axis passing under the pole at $\Omega=0$. Eq.(\ref{eq1}) can always be numerically solved by means of fast Fourier transform (FFT) but, generally, has no analytical solution. Fortunately enough, such a solution exists in the reference case of a medium with a single Lorentzian absorption-line (see, e.g., \cite{varo86}). On resonance and for \emph{  large optical thickness}, $a(t)$ takes the simple form
\begin{equation}
a(t\geq0)=\mathrm{e}^{-\gamma t}J_{0}\left(\sqrt{2\alpha L\gamma t}\right)\label{eq2}
\end{equation}
where $L$ is the medium thickness, $t$ (as in all the following)
is a \emph{local time} (real time minus $L/c$), $\alpha$ is the
resonant absorption-coefficient for the intensity ($\alpha/2$ for
the amplitude) and $\gamma$ is the half width at half maximum of
the line. For $t\geq t_{1}=\frac{1}{2\alpha L\gamma}$, the asymptotic form of $J_{0}$ may be used and $a(t)$ approximately reads
\begin{equation}
a(t\geq t_{1})\approx\sqrt{\frac{2}{\pi}}\mathrm{e}^{-\gamma t}\frac{\cos\left(\sqrt{2\alpha L\gamma t}-\pi/4\right)}{\left(2\alpha L\gamma t\right)^{1/4}}\label{eq2bis}
\end{equation}
Experimentally evidenced in \cite{bs87}, the transient given by Eq.(\ref{eq2}) and Eq.(\ref{eq2bis}) may be formally analyzed in terms of Sommerfeld and Brillouin precursors, which are temporally superimposed in dilute media \cite{jeon06,lef08}.
However we remark that these \textquotedblleft{}precursors\textquotedblright{}
precede nothing since the medium is then opaque for the \textquotedblleft{}main
field\textquotedblright{}. In order to obtain true precursors we examine
in this letter the much richer case where the medium is (nearly) transparent
at $\omega_{C}$. Our main purpose is to establish aproximate analytical expressions of the step response of such media, FFT being used to check the validity of the approximations.  

\begin{figure}[h]
\begin{centering}
\includegraphics[width=8.5cm]{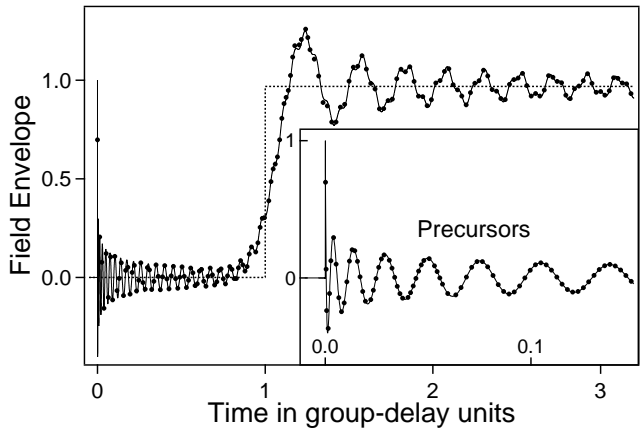} 
\par\end{centering}
\caption{Step-response $a(t)$ of a medium with a natural transparency window. The analytical $\left(\bullet\right)$ and numerical (full line) forms are respectively obtained by asymptotic calculations (see text) and by the means of a FFT involving $2^{23}$ points with a time resolution of $1.2\cdot10^{-5}\tau_{g}$. The step of amplitude $H(0)$ retarded by $\tau_{g}$ is given for reference (dotted line). Inset : enlargement of the precursors. The parameters are $\Delta=28.9\:\mathrm{ns}^{-1}$, $\gamma=0.0164\:\mathrm{ns}^{-1}$ and $\alpha L=2\cdot10^{5}$, leading to $\tau_{g}=1.96\:\mathrm{ns}$, $H(0)=\exp(-\gamma\tau_{g})=0.968$ and $b=5.22\:\mathrm{ns}^{-1}$.  \label{fig:StepResponse}}
\end{figure}

We consider first a medium with a natural transparency window between
two identical absorption-lines of intensity optical thickness $\alpha L/2\gg1$
located at $\omega_{C}\pm\Delta$. Such a medium has proved to be
a very efficient slow-light system \cite{tana03,cama06,cama07,aku08}.
Its system function reads \cite{bm06,re2}
\begin{equation}
H(\Omega)=\exp\left\{ -\frac{\alpha L\gamma}{4}\left[\frac{1}{\gamma+i\left(\Omega+\Delta\right)}+\frac{1}{\gamma+i\left(\Omega-\Delta\right)}\right] \right\} \label{eq3}
\end{equation}
A good transparency at $\Omega=0$ is achieved if $\gamma\ll\Delta$
and $\alpha L\gamma^{2}/\Delta^{2}<1$. The group delay then reads
$\tau_{g}=\alpha L\gamma/2\Delta^{2}$ \cite{bm06} and $H(0)=\exp\left(-\alpha L\gamma^{2}/2\Delta^{2}\right)=\exp\left(-\gamma\tau_{g}\right)$.
Fig.\ref{fig:StepResponse} shows the step response obtained for parameters
representative of the slow-light experiments achieved on a cesium
vapor in the near infrared \cite{cama07}.
The analytical form is obtained by taking advantage of the large value of $\alpha L$. We note first that, in its very far wings, $H(\Omega)$ equals the system function of a medium with a single line of intensity optical thickness $\alpha L$ and, as expected, the short time behavior of $a(t)$ is well described 
by Eq.(\ref{eq2}). For $t\geq t_{1}=\frac{1}{2\alpha L\gamma}=\frac{1}{4\Delta^{2}\tau_{g}}$,
$a(t)$ can be entirely calculated by the saddle point method \cite{ble86,va88}. The significant contributions to $a(t)$ originates in the relevant saddle points and, eventually, in the pole at $\Omega=0$. Introducing the phase function $\Psi(\Omega)=i\Omega t+\ln\left[H\left(\Omega\right)\right]$, Eq.(\ref{eq1})) reads
\begin{equation}
a(t)=\int_{\Gamma}\exp\left[\Psi\left(\Omega\right)\right]d\Omega/2i\pi\Omega\label{eq4}
\end{equation}
The integral is calculated by deforming $\Gamma$ in a contour $\Gamma'$
traveling along lines of steepest descent of the function $\Psi(\Omega)$
from the saddle points where $\Psi'(\Omega)=0$ . The contribution
of a non degenerate saddle point at $\Omega_{S}$ to the integral reads
\begin{equation}
a_{S}(t)=\left(i\Omega_{S}\sqrt{2\pi\left|\Psi''(\Omega_{S})\right|}\right)^{-1}\exp\left[\Psi\left(\Omega_{S}\right)+i\theta_{S}\right]\label{eq5}
\end{equation}
where $\theta_{S}$ is the angle of the direction of steepest descent
with the real axis. Note that the instantaneous frequency of $a_{S}(t)$,
defined as $d\left(\mathrm{Im}\Psi\right)/dt$ , equals $\mathrm{Re}\left(\Omega_{S}\right)$.

In the present problem, the equation $\Psi'(\Omega)=0$ giving the
saddle points can be reduced to a biquadratic equation with exact
analytic solutions. The latter can be regrouped in two pairs $\Omega_{n}^{\pm}(t)=i\gamma\pm\Omega_{n}\left(t\right)$
with $n=1,2$ and
\begin{equation}
\Omega_{n}(t)=\Delta\sqrt{1+\left[1-\left(-1\right)^{n}\sqrt{1+8t/\tau_{g}}\right]\frac{\tau_{g}}{2t}}\label{eq6}
\end{equation}
At every time, $\Omega_{1}\left(t\right)$ is real and very large compared
to $\gamma$, decreasing from $\Delta\sqrt{\tau_{g}/t}$ for $t\ll\tau_{g}$
to $\Delta$ for $t\rightarrow\infty$. The corresponding saddle points are always non-degenerate and their contribution $a_{1}\left(t\right)=a_{1}^{+}\left(t\right)+a_{1}^{-}\left(t\right)$
to $a(t)$ is easily derived from Eq.(\ref{eq5}) with $\theta_{1}^{\pm}=\pm\pi/4$.
It reads
\begin{equation}
a_{1}(t)\approx\sqrt{\frac{2}{\pi}}\mathrm{e}^{-\gamma t}\frac{\cos\left\{ \Omega_{1}t+\Delta^{2}\tau_{g}\Omega_{1}/\left(\Omega_{1}^{2}-\Delta^{2}\right)-\pi/4\right\} }{\Omega_{1}\Delta\sqrt{\tau_{g}\left[\left(\Omega_{1}+\Delta\right)^{-3}+\left(\Omega_{1}-\Delta\right)^{-3}\right]}}\label{eq7}
\end{equation}
As expected, $a_{1}(t)$ tends to $a(t)$ given by Eq.(\ref{eq2bis}) when $t\ll\tau_{g}$. More generally, $\Omega_{2}$ is purely imaginary for $t<\tau_{g}$ and the contribution of the corresponding saddle points is negligible, except in the vicinity of $\tau_{g}$. So, in a wide time-domain, $a_{1}(t)$ is actually the only significant contribution
to $a(t)$. The corresponding optical field reads $E_{1}(t)=E_{1}^{+}(t)+E_{1}^{-}(t)$
where $E_{1}^{\pm}(t)=\mathrm{Re}\left[a_{1}^{\pm}(t)\exp\left(i\omega_{C} t\right)\right]$
have instantaneous frequencies $\omega_{1}^{\pm}\left(t\right)=\omega_{C}\pm\Omega_{1}\left(t\right)$.
Due to the time dependence of these frequencies, $E_{1}^{+}(t)$ and
$E_{1}^{-}(t)$ may be identified respectively to the Sommerfeld precursor
and to the Brillouin precursor \cite{lef08}. The rise of $a\left(t\right)$ around $t=\tau_{g}$
originates from the saddle points at $\Omega_{2}^{\pm}$, which are then quasi degenerate and located in the vicinity of the pole at $\Omega =0$. The calculation of the contribution $a_{d}(t)$ to $a(t)$ of these three points requires to use an uniform asymptotic method \cite{ble86}. It is convenient to determine $a_{d}(t)$ through
the corresponding contribution $h_{d}(t)$ to the impulse response
$h\left(t\right)=\int_{-\infty}^{+\infty}\exp\left[\Psi\left(\Omega\right)\right]d\Omega/2\pi$.
Following the procedure of \cite{ble86,va88}, we get $h_{d}(t)\approx b\mathrm{e}^{-\gamma t}\mathrm{Ai}\left[-b\left(t-\tau_{g}\right)\right]$
where $\mathrm{Ai}\left(x\right)$ is the Airy function and $b=\left(\Delta^{2}/3\tau_{g}\right)^{1/3}$.
Finally $a_{d}(t)=\int_{-\infty}^{t}h_{d}(x)dx$ reads
\begin{multline}
a_{d}(t)=\mathrm{e}^{-\gamma\tau_{g}}\int_{-\infty}^{b(t-\tau_{g})}\mathrm{e}^{-\gamma x/b}\mathrm{Ai}\left(-x\right)dx
\\ \approx\mathrm{e}^{-\gamma t}\int_{-\infty}^{b(t-\tau_{g})}\mathrm{Ai}\left(-x\right)dx\label{eq8}
\end{multline}
the $2^{nd}$ form holding when $\gamma\ll b$ \cite{sha08}. $a_{d}(t)$
attains its absolute maximum at the first zero of $\mathrm{Ai}\left(-x\right)$
, that is for $x\approx2.3$ or $t=t_{2}=\tau_{g}+2.3/b$. For $t_{1}<t<t_{2}$,
$a(t)$ is well fitted by $a_{1}(t)+a_{d}(t)$ (Fig.\ref{fig:StepResponse}). For $t>t_{2}$, $\Omega_{2}$ is real and the frequencies $\Omega_{2}^{\pm}$ are well separated ($\Omega_{2}\gg\gamma$). The contribution $a_{2}(t)$ of the two saddle points to $a(t)$ can then again be derived from Eq.(\ref{eq5}) with $\theta_{2}^{\pm}=\mp\pi/4$.
It reads 
\begin{equation}
a_{2}(t)\approx-\sqrt{\frac{2}{\pi}}\mathrm{e}^{-\gamma t}\frac{\cos\left\{ \Omega_{2}t+\Delta^{2}\tau_{g}\Omega_{2}/\left(\Omega_{2}^{2}-\Delta^{2}\right)+\pi/4\right\} }{\Omega_{2}\Delta\sqrt{-\tau_{g}\left[\left(\Omega_{2}+\Delta\right)^{-3}+\left(\Omega_{2}-\Delta\right)^{-3}\right]}}\label{eq9}
\end{equation}
 The steepest descent contour $\Gamma'$ passing through the four saddle
points is now such that $\Gamma+\Gamma'$ encircles the pole in $\Omega=0$. The contributions $a_{1}(t)$ and $a_{2}(t)$ should then be completed by the corresponding residue, namely $H(0)=\mathrm{e}^{-\gamma\tau_{g}}$.
For $t>t_{2}$, we get thus $a(t)=\mathrm{e}^{-\gamma\tau_{g}}+a_{1}(t)+a_{2}(t)$. Again the agreement with the exact result is very good (Fig.\ref{fig:StepResponse}). The optical fields associated with $a_{2}(t)$ may be considered as second precursors but, since they arrive after the rise of the main field , we suggest to name them \emph{postcursors}. Contrary to those
of the precursors, their instantaneous frequencies $\omega_{2}^{\pm}\left(t\right)=\omega_{C}\pm\Omega_{2}\left(t\right)$
are initially close to $\omega_{C}$ before deviating from this frequency. Note that the oscillations in the falling tail of the pulses, observed in the experiments \cite{cama07}, are clearly related to our postcursors.

\begin{figure}[h]
\begin{centering}
\includegraphics[width=80mm]{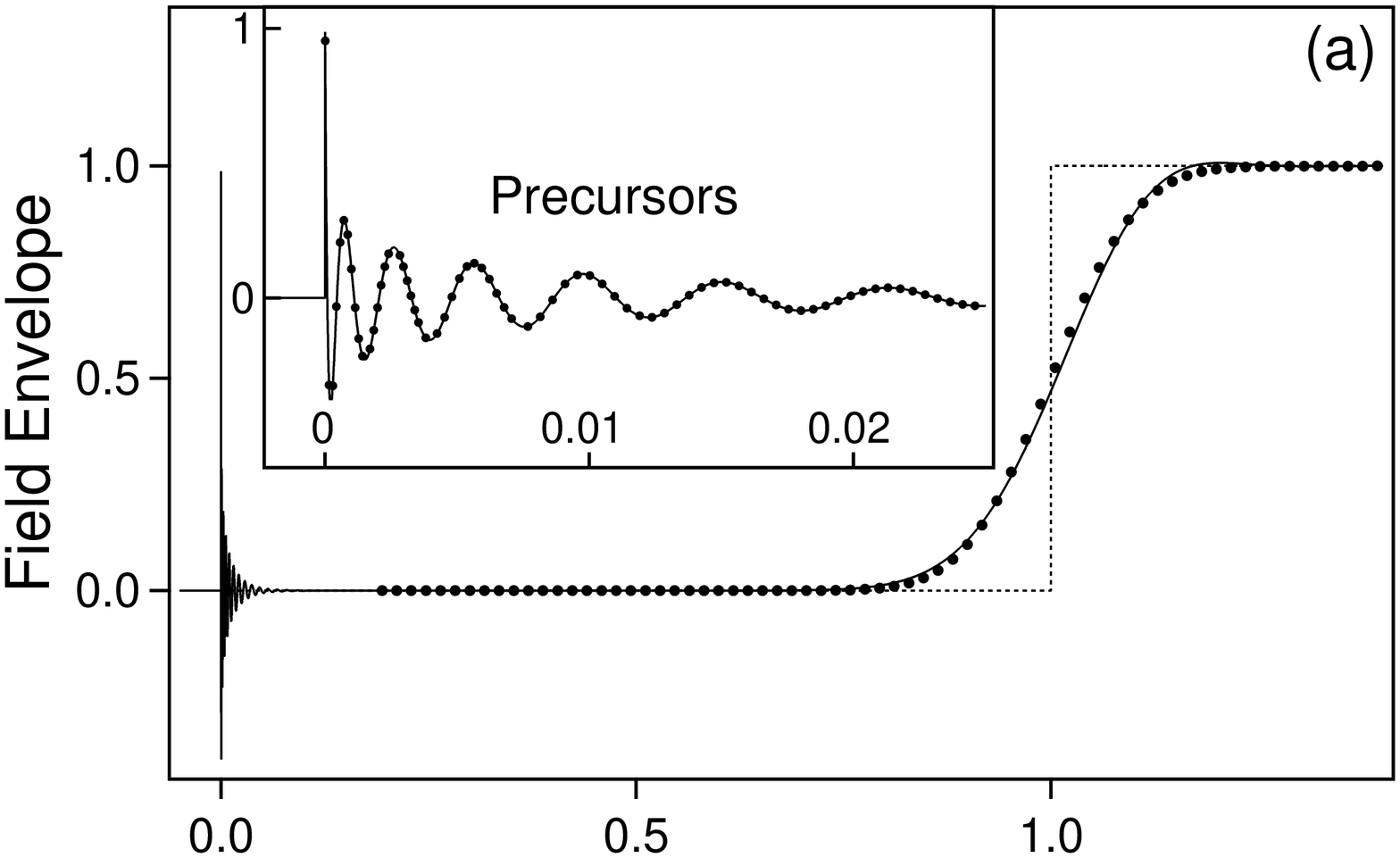}\\
\par\end{centering}
\begin{centering}
\includegraphics[width=80mm]{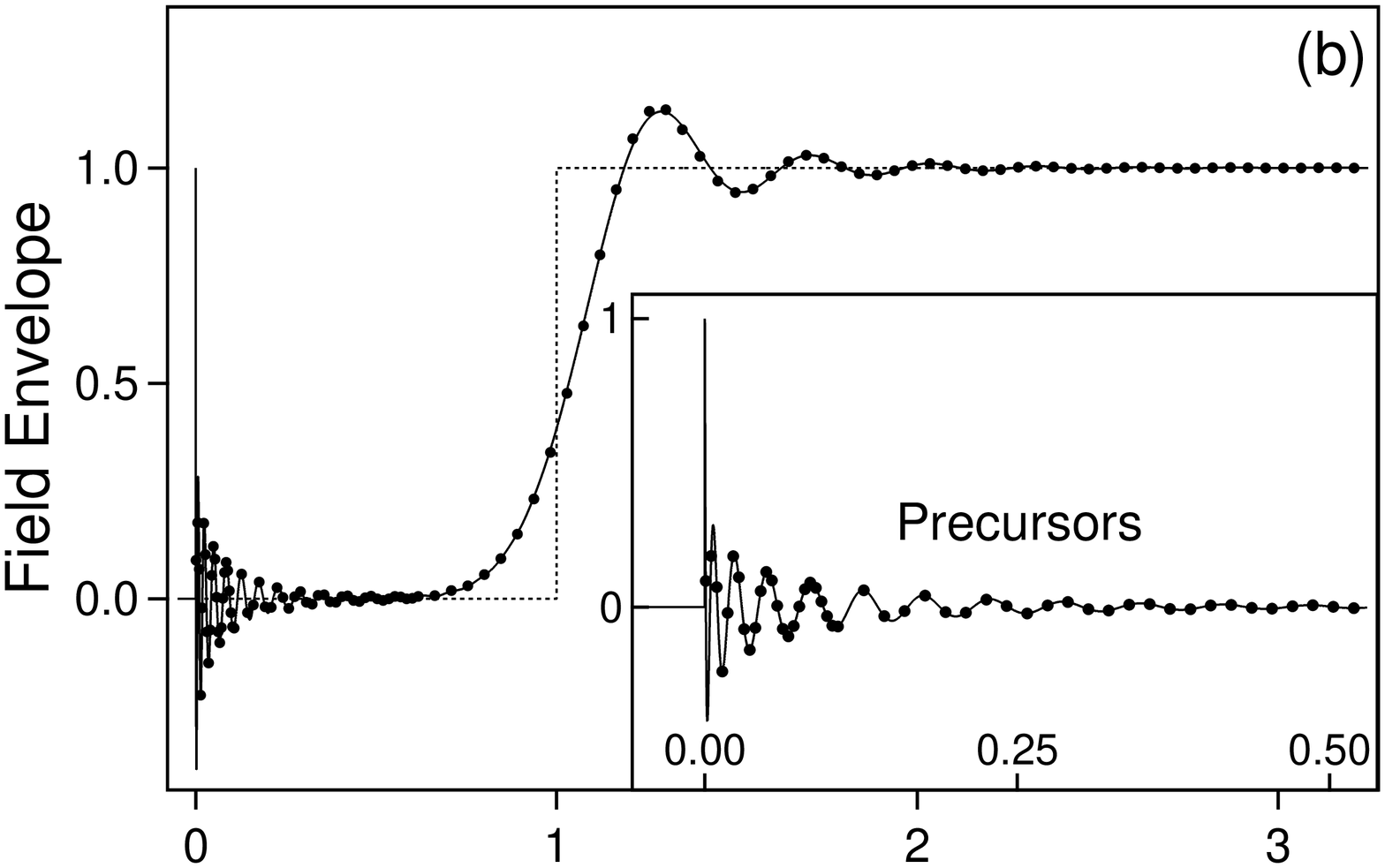}\\
\par\end{centering}
\begin{centering}
\includegraphics[width=80mm]{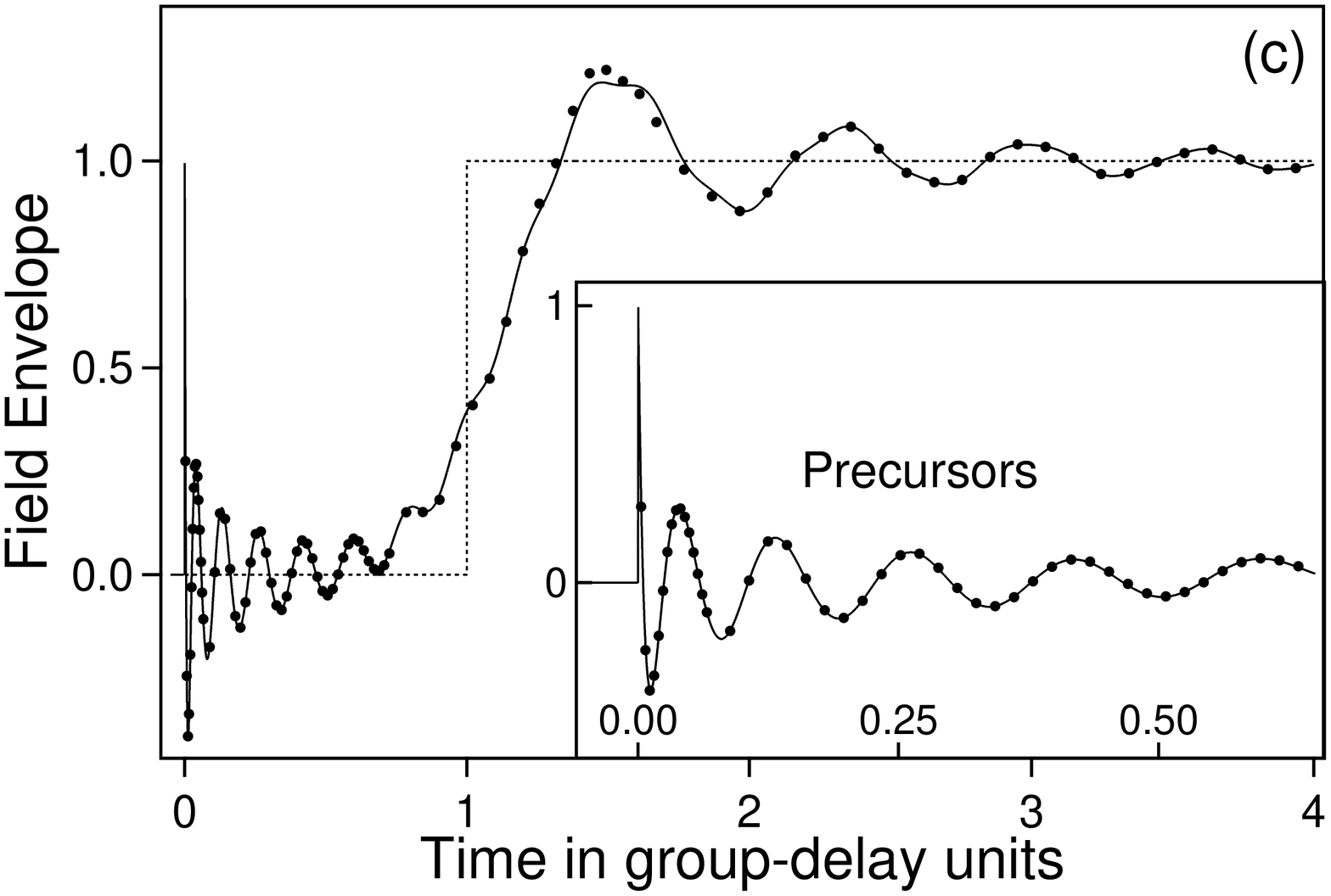} 
\par\end{centering}
\caption{Same as Fig.\ref{fig:StepResponse} for a medium with an electromagnetically
induced transparency window. The parameters are $\alpha L=600$ and
$\Omega_{r}/\gamma=$ (a) $4.60$ (b) $14.0$ (c) $34.6$, leading
to (a) $\gamma\tau_{g}=56.7$ (b) $\gamma\tau_{g}=6.12$ and $b=1.39\gamma$
(c) $\gamma\tau_{g}=1.00$ and $b=4.64\gamma$. Note that the group delays and thus the absolute time-scales are several orders larger than in the case of the natural frequency window. \label{fig:Step2}}
\end{figure}

We will now examine more briefly the case of a medium with an electromagnetically
induced transparency (EIT) window \cite{kasa95,hau99,boyd02}. In such a medium, precursors
have been indirectly demonstrated in an experiment of two-photon coincidence \cite{du08}. We consider the simplest $\Lambda$ arrangement with a resonant control field. If the coherence relaxation rate for the forbidden transition is small enough, the medium may
be transparent at $\omega_{C}$ and its system function reads
\begin{equation}
H\left(\Omega\right)=\exp\left\{ \frac{-\alpha L\gamma/2}{i\Omega+\gamma+\Omega_{r}^{2}/4i\Omega}\right\} \label{eq10}
\end{equation}
where $\Omega_{r}$ is the modulus of the Rabi frequency of the coupling
field \cite{boyd02,bm06}. We get then $\tau_{g}=2\alpha L\gamma/\Omega_{r}^{2}$.
Fig.\ref{fig:Step2} shows the step responses $a\left(t\right)$ obtained
for different $\Omega_{r}$ and for a value of $\alpha L$ intermediate
between those of the celebrated experiments achieved on a lead vapor
\cite{kasa95} and on an ultracold gas of atomic sodium \cite{hau99}. 
As previously and for the same reasons, the very short term behavior
of $a\left(t\right)$ (up to $t_{1}=\frac{1}{2\alpha L\gamma}=\frac{1}{\Omega_{r}^{2}\tau_{g}}$)
is given by Eq.(\ref{eq2}). In general, the $4^{th}$ degree equation
giving the saddle point frequencies has no simple solutions but the
following properties are easily demonstrated. Irrespective of $\Omega_{r}$,
$\Omega_{2}^{-}(\tau_{g})=0$ and, for $t\rightarrow0$, $\Omega_{1}^{\pm}\left(t\right)\rightarrow i\gamma\pm\Omega_{r}\sqrt{\tau_{g}/4t}$
while $\Omega_{2}^{\pm}\left(t\right)\rightarrow\pm i\Omega_{r}/2$.
When $\Omega_{r}<\gamma$, $\Omega_{2}^{+}\left(t\right)$ and $\Omega_{2}^{-}\left(t\right)$
keep non degenerate and purely imaginary at every time. If on the
contrary $\Omega_{r}>\gamma$, these two frequencies coalesce at a
time $t_{d}>\tau_{g}$ in $\Omega_{d}=i\Omega_{r}\sin\left[\sin^{-1}\left(\gamma/\Omega_{r}\right)/3\right]$.
For $\Omega_{r}>4\gamma$, $\Omega_{d}\approx i\gamma/3$ and $t_{d}=\tau_{g}\left(1+4\gamma^{2}/3\Omega_{r}^{2}\right)$.
Explicit analytical expressions of $a(t)$ can be obtained when $\Omega_{r}/\gamma$ is moderate or large.

In the first case, $\gamma\tau_{g}=2\alpha L\gamma^{2}/\Omega_{r}^{2}\gg1$ and
the precursors will have a short duration compared to $\tau_{g}$.
In this time domain $\Omega_{1}^{\pm}\left(t\right)\approx i\gamma t\pm\Omega_{r}\left(1+3t/2\tau_{g}\right)\sqrt{\tau_{g}/4t}$ and
\begin{equation}
a_{1}(t)\approx\sqrt{\frac{2}{\pi}}\mathrm{e}^{-\gamma t}\frac{\cos\left[\Omega_{r}\left(1+t/2\tau_{g}\right)\sqrt{t\tau_{g}}-\pi/4\right]}{\left(\Omega_{r}\sqrt{t\tau_{g}}\right)^{1/2}}\label{eq11}
\end{equation}
 If $\gamma\tau_{g}$ is extremely large, the term $t/2\tau_{g}$ may be neglected and $a_{1}(t)$ again equals $a(t)$ given Eq.(\ref{eq2bis}). This particular case is examined in \cite{du09}. When $\Omega_{r}\leq \gamma $ or when $\Omega_{r}> \gamma$ with $\gamma\left(t_{d}-\tau_{g}\right)\gg1$
(Fig.\ref{fig:Step2}a), the only other significant contribution to
$a\left(t\right)$ is $a_{2}^{-}\left(t\right)$ associated with the
saddle point at $\Omega_{2}^{-}\left(t\right)$ which tends to $0$
for $t\rightarrow\tau_{g}$. We circumvent the difficulty due to the
coincidence of the saddle point with a pole by passing through the
associated impulse response $h_{2}^{-}\left(t\right)$. It reads $h_{2}^{-}\left(t\right)=\left(\sqrt{2\pi\left|\Psi"\left(\Omega_{2}^{-}\right)\right|}\right)^{-1}\exp\left[\Psi\left(\Omega_{2}^{-}\right)+i\theta_{2}^{-}\right]$
with $\theta_{2}^{-}=0$, $\Psi\left(\Omega_{2}^{-}\right)\approx-\left[\Omega_{r}\left(t-\tau_{g}\right)/4\sqrt{\gamma\tau_{g}}\right]^{2}$
and $\Psi"\left(\Omega_{2}^{-}\right)\approx-8\gamma\tau_{g}/\Omega_{r}^{2}$.
We finally get
\begin{equation}
a_{2}^{-}\left(t\right)=\frac{1}{2}\left(1+\mathrm{erf}\left[\Omega_{r}\left(t-\tau_{g}\right)/4\sqrt{\gamma\tau_{g}}\right]\right)\label{eq12}
\end{equation}
where $\mathrm{erf}(x)$ is the error function. $a_{2}^{-}\left(t\right)\rightarrow1$
when $\Omega_{r}\left(t-\tau_{g}\right)\gg4\sqrt{\gamma\tau_{g}}$
and $a_{1}\left(t\right)+a_{2}^{-}\left(t\right)$ provides a good
approximation of the exact step response at every time (Fig.\ref{fig:Step2}a).

When $\Omega_{r}\gg\gamma$ the coupling field splits the original
line in a doublet of lines approximately centered at $\omega_{C}\pm\Omega_{r}/2$.
If, in addition, $\left(\Omega_{r}/\gamma\right)^{4}\gg8\alpha L/3$
, then $\gamma\left(t_{d}-\tau_{g}\right)\ll1$ and the situation
is analogous (but not identical) to that encountered with a natural
transparency window. The frequencies of the saddle points approximately
equal $\Omega_{n}^{\pm}\approx i\gamma_{n}\left(t\right)\pm\Omega_{n}\left(t\right)$
where $\gamma_{n}\left(t\right)\approx\left(\gamma/2\right)\left[1-\left(-1\right)^{n}\left(1+8t/\tau_{g}\right)^{-1/2}\right]$ and where $\Omega_{n}\left(t\right)$ is given by Eq.(\ref{eq6}), with
$\Delta=\Omega_{r}/2$. The different contributions to $a\left(t\right)$
then read
\begin{equation}
a_{1}(t)\approx\sqrt{\frac{2}{\left|\pi\Psi"\left(\Omega_{1}^{+}\right)\right|}}\mathrm{Re}\left\{ \frac{1}{\Omega_{1}^{+}} \mathrm{e}^{\left[\Psi\left(\Omega_{1}^{+}\right)-i\pi/4\right]}\right\} \label{eq13}
\end{equation}
\begin{equation}
a_{2}(t)\approx -\sqrt{\frac{2}{\left|\pi\Psi"\left(\Omega_{2}^{+}\right)\right|}}\mathrm{Re}\left\{ \frac{1}{\Omega_{2}^{+}} \mathrm{e}^{\left[\Psi\left(\Omega_{2}^{+}\right)+i\pi/4\right]}\right\}  \label{eq14}
\end{equation}
\begin{equation}
a_{d}\left(t\right)\approx\int_{-\infty}^{b\left(t-\tau_{g}\right)}\mathrm{Ai}\left(-x\right)\exp\left(-\gamma x/3b\right)dx\label{eq15}
\end{equation}
where $b=\left(\Omega_{r}^{2}/12\tau_{g}\right)^{1/3}$. As in the
case of the natural frequency window, $a_{1}\left(t\right)+a_{d}\left(t\right)$
and $a_{1}\left(t\right)+a_{2}\left(t\right)+H(0)$ fit very well
the exact step response, respectively for $t_{1}<t<t_{2}=\tau_{g}+2.3/b$
and for $t>t_{2}$ (Fig.\ref{fig:Step2}b and Fig.\ref{fig:Step2}c).
The main difference is that a significant damping of the precursors
is now compatible with a good transparency at $\omega_{C}$. For intermediate
values of $\Omega_{r}$ it is so possible to observe both well developed
precursors and postcursors without overlapping (Fig.\ref{fig:Step2}b).
On the contrary, the tail of the precursors again partially interferes
with the postcursors for very large $\Omega_{r}$(Fig.\ref{fig:Step2}c).

To conclude, we have obtained, for the first time, fully analytic expressions of the entirety of the 
step response of linear media with a transparency window. Our results
show that these media, contrary to those generally considered, are
well adapted to observe in a same experiment the precursors, the main
field and the postcursors, all well distinguishable from each other
and having comparable amplitudes. Insofar as the parameters used in
the calculations are representative of real experiments, we think
that our work might stimulate an experimental observation of these
rich dynamics, which would, in turn, stimulate new theoretical investigations
on related slow-light systems.


\begin{thebibliography}{10}
\bibitem{re2} The results derived from Eq.(3) hold even if the two lines are not Lorentzian provided that they are so in their wings \cite{cama07}. The optical thickness $\alpha L/2$ intervening in this equation may then be considerably larger than its actual value on resonance.
\end{thebibliography}
\end{document}